\documentclass[superscriptaddress,nofootinbib,10pt,preprintnumbers]{revtex4}
\usepackage{graphicx,amssymb,amsmath,slashed}
\usepackage{color,graphicx,epsfig}
\usepackage{ifpdf}
\usepackage{amsmath}
\usepackage{bm}
\usepackage{color}
\usepackage[english]{babel}
\usepackage{graphicx}%
\usepackage{amsfonts}%
\usepackage{amssymb}
\usepackage{braket}
\usepackage{hyperref}

\definecolor{nicered}{rgb}{0.7,0.1,0.1}
\definecolor{nicegreen}{rgb}{0.1,0.5,0.1}
\hypersetup{colorlinks,citecolor= nicegreen,linkcolor= nicered}

\newcommand{\beq}{\begin{equation}}
\newcommand{\eeq}{\end{equation}}

\begin{document}

\title{\boldmath On the $B\to D^* \tau \bar \nu_\tau$ Sensitivity to New Physics}

\author{Svjetlana Fajfer}
\affiliation{J. Stefan Institute, Jamova 39, P. O. Box 3000,
1001 Ljubljana, Slovenia} \affiliation{Department of Physics,
University of Ljubljana, Jadranska 19, 1000 Ljubljana,
Slovenia}

\author{Jernej F.\ Kamenik}
\affiliation{J. Stefan Institute, Jamova 39, P. O. Box 3000,
1001 Ljubljana, Slovenia} \affiliation{Department of Physics,
University of Ljubljana, Jadranska 19, 1000 Ljubljana,
Slovenia}

\author{Ivan Ni\v sand\v zi\'c}
\affiliation{J. Stefan Institute, Jamova 39, P. O. Box 3000,
1001 Ljubljana, Slovenia}

\begin{abstract}
 B physics has played a prominent role in investigations of new physics effects at low-energies. 
 Presently, the largest discrepancy between a standard model prediction  and experimental measurements appears in the branching ratio of the charged current mediated $B \to \tau \bar \nu_\tau$ decay, where the large $\tau$ mass lifts the helicity suppression arising in leptonic B decays. Less significant systematic deviations are also observed in the semileptonic $B \to D^{(*)} \tau \bar \nu_\tau$ rates.  Due to the rich spin structure of the final state, the decay mode $B \to D^* \tau \bar \nu_\tau$ offers a number of tests of such possible standard model deviations.  We investigate the most general set of lowest dimensional effective operators leading to helicity suppressed modifications of $b\to c$ (semi)leptonic transitions. We explore such contributions to the  $B\to D ^* \tau \bar \nu_\tau$ decay amplitudes by  determining the differential decay rate,  longitudinal $D^*$ polarization fraction,  $D^* - \tau$ opening angle asymmetry and the $\tau$ helicity asymmetry. We identify the size of possible new physics contributions to these observables constrained by the present $B\to D^{(*)} \tau \bar \nu_\tau$  rate measurements and find significant modifications are still possible in all of them. In particular, the opening angle asymmetry can be shifted by almost   $30\%$, relative to the standard model prediction, while the $\tau$ helicity asymmetry can still deviate by as much as $80\%$. 
\end{abstract}

\maketitle

%
\section{Introduction}
%

Several recent experimental results in $B$ physics have significantly constrained the possibility of large New Physics (NP) effects in rare $B$ processes. In particular, new CP violating effects in the $B_s\to J/\psi \phi$ decay  are already constrained to be of the order of the Standard Model (SM) expectations~\cite{Lenz:2012az}. Similarly, the recent LHCb bound on $Br(B_s \to \mu^+ \mu^-)$~\cite{LHCbMoriond} already rules out any significant enhancement over the SM prediction in this decay.

On the other hand, existing  measurements of the branching ratio for the charged current mediated $B \to \tau \bar \nu_\tau$ process yield results which are systematically higher than the SM expectations~\cite{Bozek:2010zz} and the current world average for this lepton helicity suppressed decay rate is a factor of 2 above the SM predicted value~\cite{Charles:2011va}. 
B meson decays with $\tau$ leptons in the final state offer possibilities of significant NP contributions not present in processes with light leptons. Namely, the large tau mass can uplift the helicity suppression of  certain (semi)leptonic decay amplitudes which are unobservable in decays with light leptons in the final state. NP models with extended ElectroWeak (EW) symmetry breaking sectors -- the Two Higgs Doublet Models (THDMs) being the canonical examples --  often predict enhancements in such helicity suppressed amplitudes. Existing studies of the $B\to D\tau \bar \nu_\tau$ decay~\cite{Tanaka:1994ay, Nierste:2008qe, Kamenik:2008tj, Tanaka:2010se} have already shown how such NP effects can be over constrained, and how additional kinematical observables in the three body decay offer tests of the underlying short distance contributions not possible in the two body leptonic mode.  

In this respect, the $B \to D^* \tau \bar \nu_\tau$ decay having two detectible particles of non-zero spin in the final state ($D^*, \tau$) offers the opportunity of an even more complete investigation of the structure of possible NP contributions to $b\to c \tau \bar \nu_\tau$ transitions~\cite{Tanaka:1994ay}. The experimental reconstruction of the $D^*$ in the  $D \pi$ final state allows to obtain the helicity structure of this state directly. Similarly, the  $\tau$ lepton helicity can be inferred from its decays to $\pi \nu_\tau$ final states~\cite{Nierste:2008qe, Tanaka:2010se}. This means that a number of experimental observables sensitive to possible NP effects, can be introduced.  In the present study, we explore several such observables, like the differential distribution over the lepton invariant mass,  the longitudinal $D^*$ branching fraction, a $D^* -\tau$ opening angle asymmetry, as well as the $\tau$ helicity fractions. To this end we employ a model independent effective field theory approach and identify NP contributions, which naturally predict helicity suppressed contributions in (semi)leptonic B meson decays while preserving the well established SM form of charged lepton currents (c.f.~\cite{Dassinger:2008as} for a recent related discussion). 

The paper is organized as follows: in Sec.~\ref{sec:NP} we investigate NP inducing helicity suppressed contributions to  $b\to c \ell \bar \nu_\ell$  within the effective theory approach and evaluate existing constraints coming from the $B\to D\tau \bar \nu_\tau$ rate measurements. In Sec.~\ref{sec:Rates} we focus on the various kinematical distributions and spin observables accessible in the $B \to D^* \tau \bar \nu_\tau$ mode and estimate their sensitivity to such NP, while the explicit derivation of the relevant polarized differential rates and evaluation of the corresponding hadronic matrix elements is relegated to the Appendices.  Finally, we conclude in Sec.~\ref{sec:Conclusions}.

%
\section{New Physics in $\bf b\to c \tau \bar \nu_\tau$}
\label{sec:NP}
%

Following~\cite{Dassinger:2008as}, we consider effective weak Hamiltonian, relevant for $b\to c \ell\bar\nu_\ell$ transitions in presence of NP contributing only to charged current interactions of quarks, while manifiestly preserving the well tested universal $V-A$ structure of leptonic charged currents
\begin{equation}
\begin{split}
\mathcal{H}_{\rm eff}=&\frac{4 G_F V_{cb}}{\sqrt{2}} J_{bc,\mu} \sum_{\ell=e,\mu,\tau} \left( \bar \ell \gamma^\mu P_L  \nu_\ell  \right) +\rm h.c.\,,\\
\end{split}
\label{eq:Heff}
\end{equation}
where $P_{L,R}\equiv (1\mp \gamma_5)/2$, while $J_{bc}^\mu$ is the effective $b\to c$ charged current. In particular, we are interested in NP contributions, which lead to charged lepton helicity suppression in $B\to D^{(*)}\ell \bar \nu_\ell$, and are thus inaccessible in semileptonic decays to light leptons~\cite{Dassinger:2008as}.
In general, this is the case if the NP contributions to $J_{bc}^\mu$ can be written as a total derivative of a scalar operator. In the effective field theory expansion, the most relevant (lowest dimensional) contributions to $J_{bc}^\mu$ are then
\beq
J_{bc}^\mu = \bar c \gamma^\mu P_L b + g_{SL} i \partial^\mu (\bar c P_L b ) + g_{SR} i \partial^\mu (\bar c P_R b )
\,,
\label{eq:Jeff}
\eeq
where the first term corresponds to the SM charged current, while $g_{SL,SR}$ are dimensionful NP couplings. If the NP contributions are associated with a high NP scale $\Lambda_{\rm NP}\gg v_{\rm EW}$, then $g_{SL,SR} \sim 1/\Lambda_{\rm NP}$. 
A particular and well known realization of such NP contributions is the THDM type II where only $g_{SR}$ receives a significant contribution. It is of the form $g_{SR} \sim - m_b \tan^2\beta/m_{H^+}^2$~\cite{thdm2} where $\tan\beta$ is the ratio of the two EW condensates in the model, and $m_{H^+}$ is the mass of the physical charged Higgs boson. 
Further NP contributions to $J_{bc}^\mu$ relevant for helicity suppressed decays can be obtained via insertions of the $\partial^2$ operator and are thus necessarily suppressed by at least two additional powers of $\Lambda^{-1}_{\rm NP}$.  

In specific models one may relate NP effects in $b\to c$ transitions to other sectors, i.e. $b,s \to u$ ($B\to \tau\bar \nu_\tau$, $K\to \mu\bar \nu_\mu$) or $c\to d,s$ ($D_{(s)} \to \ell \bar \nu_\ell$), c.f.~\cite{Deschamps:2009rh, Dorsner:2009cu}, resulting in a more constrained parameter space. In the present study we will however not assume any underlying flavor structure and focus exclusively on observables in the $b\to c$ sector. Before exploring such NP effects in various kinematical distributions of the $B\to D^*\tau\bar\nu_\tau$, we need to consider existing constraints coming from the measurement of the $B\to D\tau\bar \nu_\tau$ decay rate. In particular it turns out that most hadronic and SM parametric uncertainties cancel in the ratio between the tau and light lepton branching ratios~\cite{Chen:2006nua, Kamenik:2008tj}, i.e.
\beq
R\equiv \frac{Br(B\to D \tau \bar \nu_\tau)}{Br(B\to D e\bar \nu_e)}\,.
\eeq
This ratio can already be predicted with considerable precision in the SM, and present estimates using either Lattice results with $R^{\rm Latt.}_{\rm SM}=0.296(16)$~(updated value based on~\cite{Kamenik:2008tj} using the recent precise form factor shape determination from~\cite{HFAG}) or heavy quark expansion with $R^{\rm HQET}_{\rm SM}=0.302(15)$~\cite{Tanaka:2010se} agree within the stated errors. NP of the form in~\eqref{eq:Jeff} results in a modification of the $R$ ratio between the tau and light lepton rates
\beq
R/R_{\rm SM} = 1 + 1.5 {\rm Re}[m_\tau (g_{SR}+g_{SL})] +1.0|m_\tau (g_{SR}+g_{SL})|^2\,,
\eeq
where we have again updated the expression in~\cite{Kamenik:2008tj} using the form factor shape determination from~\cite{HFAG}, and the $\overline{\rm MS}$ values at the $m_B$ scale have been used for the bottom and charm quark masses. Comparing these expressions with the experimentally determined values~\cite{PDG}
\begin{align}
Br(B^+\to \bar D^0 \tau^+ \nu_\tau)_{\rm exp} &= (0.77\pm 0.25)\%\,,\nonumber\\
Br(B^+\to \bar D^0 \ell^+ \nu_\ell)_{\rm exp} &= (2.23\pm 0.11)\%\,,\quad {\rm for}~\ell = e, \mu\,,
\end{align}
one can obtain constraints~\footnote{The related neutral $B$ decay modes have also been measured~\cite{PDG} in both $D$ and $D^*$ final states, however with less significance compared to the charged $B$ modes resulting in less stringent bounds.} in the complex plane of $(g_{SR}+g_{SL})$ as shown in the left plot in Fig.~\ref{fig:Rbound}.

For completeness, we provide SM predictions for the branching fractions by using the experimentally measured decay rates to light leptons, i.e. assuming no NP in those modes. Using inputs from~\cite{PDG, HFAG} we obtain
\begin{align}
Br(B^+\to \bar D^0 \tau^+ \nu_\tau)_{\rm SM} &= (0.66\pm0.05)\%\,, & Br(B^0\to D^- \tau^+ \nu_\tau)_{\rm SM} &= (0.64\pm0.05)\%\,.
\end{align}

%
\section{$B \to D^* \tau \nu_\tau$ differential decay rates}
\label{sec:Rates}
%

We consider the decay of a $B$ meson to a polarized $D^*$ (of helicity $+,-$ or $0$), a $\tau$ lepton of a given helicity ($\lambda_\tau = \pm1/2$), and $\bar \nu_\tau$ (with helicity $\lambda_\nu = 1/2$) as mediated by the effective Hamiltonian of the form in Eq.~\eqref{eq:Heff}.  The relevant kinematical variables describing the three-body decay are $q^2 \equiv (p_B-p_{D^*})^2$, where $p_{B,D^*}$ are the $B$ and $D^*$ momenta, respectively, and the angle $\theta$ between the $D^*$ and $\tau$ three-momenta in the $\tau-\bar\nu_\tau$ restframe. The detailed derivation of the polarized double differential rates is given in the Appendix~\ref{sec:A1} with the final result in Eq.~\eqref{eq:Gammaf}.

We focus first on the decay distributions in absence of tau helicity information. Summing over both tau helicities $\lambda_{\tau}$ in~\eqref{eq:Gammaf}, we obtain
\begin{equation}
\begin{split}
\frac{d^2\Gamma_\tau}{dq^2 d \cos\theta}=&\frac{G_F^2 |V_{cb}|^2|{\bf p}\,|q^2}{256\pi^3 m_B^2} \left(1-\frac{m_\tau^2}{q^2}\right)^2\times\\
&\Bigg[(1-\cos\theta)^2 |H_{++}|^2+(1+\cos\theta)^2 |H_{--}| ^2+2\sin^2\theta |H_{00}|^2+ \\
&\frac{m_\tau^2}{q^2}\Big((\sin^2\theta(|H_{++}|^2+|H_{--}|^2)+2|H_{0t}-H_{00}\cos\theta|^2\Big)\Bigg],
\end{split}
\label{eq:dGamma}
\end{equation}
where $|{\bf p}\,|$ is defined in Eq.~\eqref{eq:pD}, and $H_{mn}$ are the relevant ($q^2$ dependent) helicity amplitudes, defined in Appendix~\ref{sec:A2}. Performing integration over $d \cos\theta$ in~\eqref{eq:dGamma}, we obtain
\begin{equation}
\frac{d\Gamma_\tau}{dq^2}=\frac{G_F^2 |V_{cb}|^2|{\bf p}\,|q^2}{96\pi^3 m_B^2} \left(1-\frac{m_\tau^2}{q^2}\right)^2\left[\left(|H_{++}|^2+|H_{--}|^2+|H_{00}|^2\right)\left(1+\frac{m_\tau^2}{2q^2}\right)+\frac{3}{2}  \frac{m_\tau^2}{q^2}|H_{0t}|^2\right],
\label{eq:dGamma1}
\end{equation}
in agreement with the well known result~\cite{Korner-Schuler1, Pham}. The presence of NP quark charged currents defined in~\eqref{eq:Jeff} only affects the $H_{0t}$ helicity amplitude and can be encoded compactly as   
\beq
H_{0t} = H_{0t}^{\rm SM} \left[1 + (g_{SR}-g_{SL}) \frac{q^2}{m_b+m_c}\right]\,.
\label{eq:HNP}
\eeq
In the numerical evaluation of such NP effects we use the $\overline{\rm MS}$ values for the bottom and charm quark masses at the $m_b$ scale. The task of extracting information on NP from the differential decay rates thus reduces to obtaining sensitivity to the $H_{0t}$ helicity amplitude. 

\begin{figure}[t]
\centering
\includegraphics[width=0.25\textwidth]{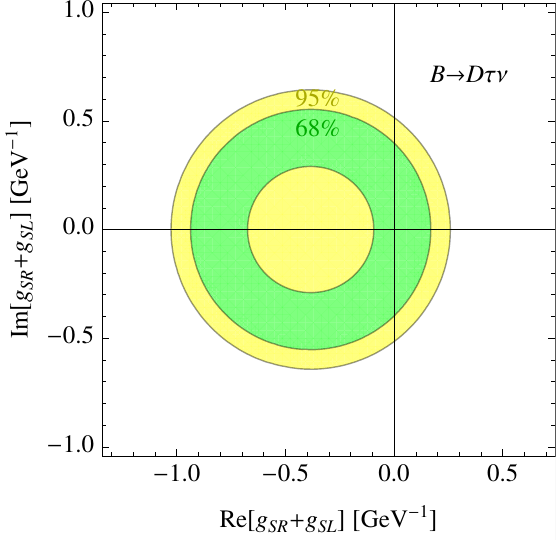}
\includegraphics[width=0.243\textwidth]{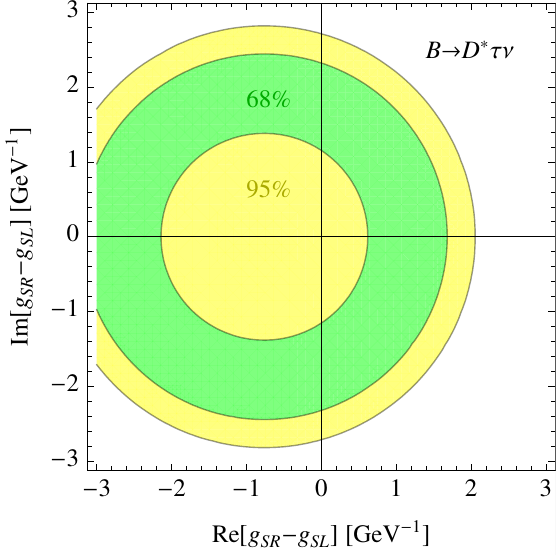}
\includegraphics[width=0.25\textwidth]{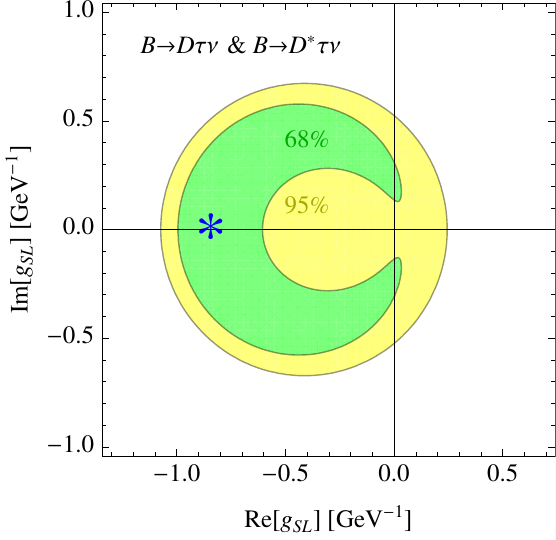}
\caption{\label{fig:Rbound} The $68\%$ (in darker green shade) and $95\%$ (in lighter yellow shade) C.L. allowed regions in the complex plane of NP parameters appearing in the effective $b\to c$ charged current in Eq.~\eqref{eq:Jeff}.  Shown are fixed combinations (the orthogonal combinations are set to zero): $g_{SR}+g_{SL}$ (left plot) bounded by $B\to D \tau \bar \nu_\tau$, $g_{SR}-g_{SL}$ (center plot) constrained only by $B\to D^* \tau \bar \nu_\tau$, and $g_{SL}$ (right plot) contributing to both modes. The best fit NP benchmark point in the later case is marked with the symbol $*$.}
\end{figure}

We start with the $B\to D^* \tau \bar \nu_\tau$ branching fractions by integrating~\eqref{eq:dGamma1} over $q^2$. These are also the only already measured observables in these modes. As in the case of the $B\to D \tau \bar \nu_\tau$ decay, most theoretical uncertainties, related to the evaluation of the hadronic form factors defined in Appendix~\ref{sec:A2} are significantly reduced if one normalizes the $B\to D^* \tau \bar \nu_\tau$ rates to the modes with the light charged leptons in the final state~\cite{Chen:2006nua} -- one considers the ratio
\beq
R^* \equiv \frac{Br(B\to D^* \tau \bar \nu_\tau)}{Br(B\to D^* e\bar \nu_e)}\,.
\eeq
In this way we obtain
\beq
R^* = R^*_{\rm SM} \left\{ 1 + 0.12 {\rm Re} [ m_\tau (g_{SR}-g_{SL}) ] + 0.05 | m_\tau (g_{SR}-g_{SL}) |^2 \right\}\,,
\eeq
where we find for the SM prediction (using the recent precise experimental extraction~\cite{Belle} of the relevant form factor ratios)\footnote{The value is obtained by averaging over the $B^\pm$ and $B^0$ modes, for which in absence of EM corrections, $R^*_{\rm SM}$ differs by less than 0.001.}
\beq
R^*_{\rm SM}  = 0.252(3)\,.
\eeq
The stated hadronic uncertainty is dominated by the estimate of higher order perturbative and power corrections to the heavy quark limit of the $A_0/A_1$ form factor ratio which presently cannot be extracted directly from data (see Appendix~\ref{sec:A2}). At this level of precision, EM corrections affecting $B\to D^* e\bar \nu_e$ and  $B\to D^* \tau\bar \nu_\tau$ differently  could become important~\cite{Becirevic:2009fy} but the related uncertainty due to such effects depends on the particular experimental setup and would require a dedicated study beyond the scope of the present paper.  The above expressions are to be compared with the experimentally determined branching fractions~\cite{PDG}
\begin{align}
Br(B^+\to \bar D^{*0} \tau^+ \nu_\tau)_{\rm exp} &= (2.1\pm 0.4)\%\,,\nonumber\\
Br(B^+\to \bar D^{*0} \ell^+ \nu_\ell)_{\rm exp} &= (5.68\pm 0.19)\%\,,\quad {\rm for}~\ell = e, \mu\,.
\end{align}
From this we can again obtain constraints in the complex plane of $(g_{SL}-g_{SR})$ as shown in the central plot in Fig.~\ref{fig:Rbound}. We observe that while certainly being complementary to the $B\to D\tau\bar\nu_\tau$ mode, NP contributions to the integrated  $B\to D^*\tau\bar\nu_\tau$  branching fraction are much more diluted. 

We also note that at present, the experimental measurements are systematically above SM predictions in $B\to D^{(*)}\tau\bar\nu_\tau$ decays. This is clearly demonstrated in the special case, where we set $g_{SR}=0$ and study the combined constraints from both decay modes in the right plot in Fig.~\ref{fig:Rbound}. We observe that such a fit mildly prefers a non-SM solution with $g_{SL} \simeq -0.9\,$GeV${^{-1}}$. In the following we will use this benchmark point to evaluate the discriminating power of the various $B\to D^*\tau\bar\nu_\tau$ observables.
 
Again for completeness, we provide SM predictions for the $B\to D^* \tau\bar \nu_\tau$ branching fractions by normalizing the value of $R^{*}_{\rm SM}$ to the experimentally measured decay rates to light leptons, i.e. assuming no NP in those modes. Using inputs from~\cite{PDG, Belle} we obtain
\begin{align}
Br(B^+\to \bar D^{*0} \tau^+ \nu_\tau)_{\rm SM} &= (1.43\pm0.05)\%\,, & Br(B^0\to D^{*-} \tau^+ \nu_\tau)_{\rm SM} &= (1.29\pm0.06)\%\,,
\end{align}
in agreement with previous estimates~\cite{Chen:2006nua}.

\begin{figure}[t]
\centering
\includegraphics [width=0.4\textwidth]{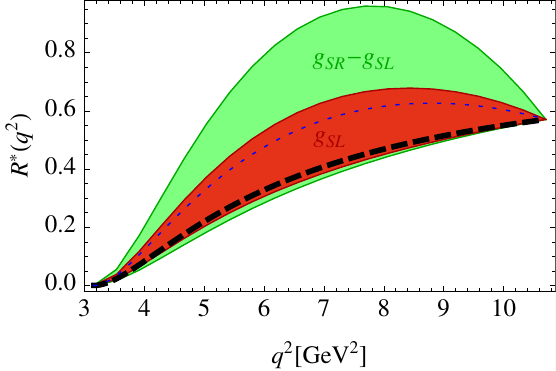}
\includegraphics [width=0.4\textwidth]{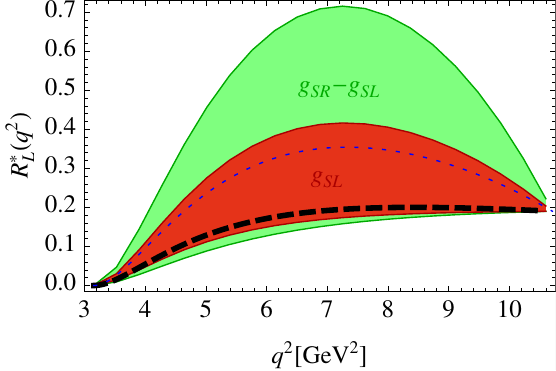}
\caption{The differential ratios $R^*(q^2)$ (left plot) and $R_L^*(q^2)$ (right plot) as functions of $q^2$. The black dashed curves are the SM predictions, while the blue dotted lines denote predictions for the NP benchmark point  (see text for details).  The $95\%$ C.L. allowed regions (due to existing constraints from the $B\to D^{(*)} \tau \bar \nu_\tau$ branching ratio measurements) for fixed NP parameter combinations (the orthogonal combinations being set to zero) $g_{SR}-g_{SL}$ and $g_{SL}$ are shown in lighter green and darker red shades, respectively. The $g_{SR}$ case is almost degenerate with $g_{SL}$ and is therefore not shown.\label{fig:Rs}}
\end{figure}

Let us next explore the influence of NP on the ratio of singly differential decay rates with tau and light leptons in the final state
\begin{equation}
R^*(q^2)=\frac{d\Gamma_{\tau}/dq^2}{d\Gamma_{\ell}/dq^2} =\left(1-\frac{m_\tau^2}{q^2}\right)^2 \left[ \left(1+\frac{m_\tau^2}{2q^2}\right)+\frac{3}{2}\frac{m_\tau^2}{q^2}\frac{|H_{0t}|^2}{|H_{++}|^2+|H_{--}|^2+|H_{00}|^2}\right]\,.
\label{eq:R}
\end{equation}
Here, $d\Gamma_{\ell}/dq^2$ is the differential decay rate to a light charged lepton, where helicity suppressed effects are negligible. Taking into account present constraints on the NP contributing in linear combinations $g_{SL}$ and $g_{SR}-g_{SL}$ we obtain the $95\%$ C.L. allowed bands in the left plot in Fig.~\ref{fig:Rs}.
We observe that significant effects in $R^*(q^2)$ (and consequently $R^*$) are still possible, especially if NP contributions are aligned with the $g_{SR}-g_{SL}$ direction -- if they appear in the form of a pseudo-scalar density operator to which $B\to D \tau \bar \nu_\tau$ has no sensitivity.

\subsection{Longitudinal $D^*$ Polarization and the Opening Angle Asymmetry}

Since NP of the form~\eqref{eq:Jeff}  only contributes to longitudinally polarized $D^*$ ($D^*_L$) in the final state, an increased sensitivity can be expected by using information on the polarization of the $D^*$, which can be inferred from the angular distributions of its decay products (i.e. $D\pi$). In~\eqref{eq:dGamma1} only $H_{00}$ and  $H_{0t}$ contribute $D^*_L$'s, leading to a prediction for the longitudinal rate, again normalized to the light lepton mode
\beq
R^*_L \equiv \frac{Br(B\to D^*_L \tau \bar \nu_\tau)}{Br(B\to D^* e \bar \nu_\tau)} = 0.115(2) \left\{ 1 + 0.27 {\rm Re} [ m_\tau (g_{SR}-g_{SL}) ] + 0.10 | m_\tau (g_{SR}-g_{SL}) |^2 \right\}\,,
\eeq
where we have also given the estimated hadronic uncertainty of the SM prediction. In addition to this inclusive observable, one can also study the singly differential longitudinal rate ratio $R^*_L(q^2)$ defined analogously to $R^*(q^2)$ in Eq.~\eqref{eq:R}. The presently allowed ranges for this observable are shown in the right plot in Fig.~\ref{fig:Rs}. Compared to $R^*(q^2)$ this observable clearly exhibits an increased sensitivity to NP contributions.
 
\begin{figure}[t]
\centering
\includegraphics[width=0.4\textwidth]{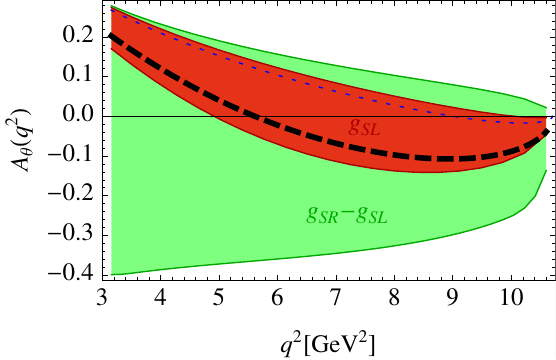}
\includegraphics [width=0.4\textwidth]{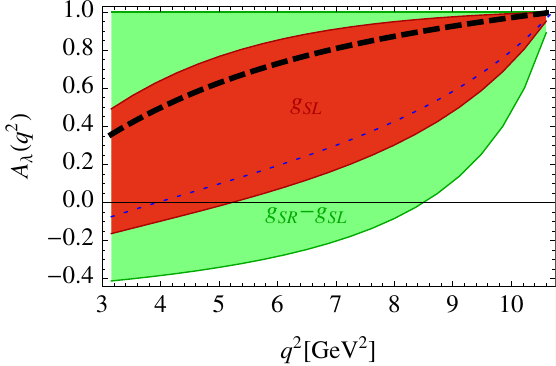}
\caption{The differential asymmetries $A_\theta(q^2)$ (left plot) and $A_\lambda(q^2)$ (right plot) as functions of $q^2$. The black dashed curves are the SM predictions, while the blue dotted lines denote predictions for the NP benchmark point (see text for details).  The $95\%$ C.L. allowed regions (due to existing constraints from the $B\to D^{(*)} \tau \bar \nu_\tau$ branching ratio measurements) for fixed NP parameter combinations (the orthogonal combinations being set to zero) $g_{SR}-g_{SL}$ and $g_{SL}$ are shown in lighter green and darker red shades, respectively. The $g_{SR}$ case is almost degenerate with $g_{SL}$ and is therefore not shown.\label{fig:Asym1}}
\end{figure}

While the $D^*$ polarization information can be extracted directly from the angular distribution in~\eqref{eq:dGamma}, this requires experimental fits to two-dimensional decay distributions. In order to avoid such challenges, we propose a simple angular (opening angle) asymmetry defined as the difference between partial rates where the angle $\theta$ between the $D^*$ and $\tau$ three-momenta in the $\tau-\bar \nu_\tau$ rest-frame is greater or smaller than $\pi/2$
\begin{align}
A_{\theta}(q^2) &\equiv \frac{\int_{-1}^{0} d\cos\theta (d^2\Gamma_\tau /dq^2 d\cos\theta) - \int_0^{1} d\cos\theta (d^2\Gamma_\tau /dq^2 d\cos\theta)}{d\Gamma_\tau/dq^2} \nonumber\\
&= \frac{3}{4}\frac{|H_{++}|^2 - |H_{--}|^2 +  2 \frac{m_\tau^2}{q^2} {\rm Re} (H_{00} H_{0t}) }{\left[\left(|H_{++}|^2+|H_{--}|^2+|H_{00}|^2\right)\left(1+\frac{m_\tau^2}{2q^2}\right)+\frac{3}{2}  \frac{m_\tau^2}{q^2}|H_{0t}|^2\right]} \,.
\end{align}
In the decay modes with light leptons, this asymmetry ($A^\ell_\theta$) can be used to probe for the presence of right-handed $b\to c$ currents, since these contribute with opposite sign to $H_{\pm\pm}$ relative to the SM. In the tau modes, it is sensitive only to the real part of NP $g_{SL}-g_{SR}$ contributions and thus provides complementary information compared to the total rate (or $R^*$). The presently allowed ranges for the $A_\theta$ asymmetry are shown in the left plot in Fig~\ref{fig:Asym1}. We observe that significant deviations from the SM prediction in this observable are still allowed. Also note that in the SM this observable exhibits a zero crossing at $q_0^2 \simeq 5.6$~GeV${}^2$\,, while this is not necessarily the case in presence of NP. 
On the other hand, the inclusive asymmetry $A_\theta$ integrated over $q^2$ is very small in the SM with $A_{\theta,\rm SM} = -6.0(8)\%$; for our NP benchmark point we obtain $A_{\theta,\rm NP} = 3.4\%$\,, but even values as low as $-30\%$ are still allowed.

\subsection{Using $\tau$ helicity }

It has been pointed out recently~\cite{Tanaka:2010se}, that the spin of the tau lepton originating from semileptonic $B$ decays can be inferred using the distinctive tau decay patterns. Therefore it is beneficial to consider the $B\to D^* \tau \bar \nu_\tau$ decays with taus in a given helicity state ($\lambda_\tau=\pm1/2$). In particular, assuming the standard $V-A$ structure of the leptonic charged current entering the relevant effective weak Hamiltonian~\eqref{eq:Heff}, the $\lambda_\tau =1/2$ helicity final states are suppressed by the tau lepton mass. Using the derivation of the polarized differential decay rates in Appendix~\ref{sec:A1} we obtain
\begin{align}
\frac{d\Gamma_\tau}{dq^2}(\lambda_\tau=-1/2)&=\frac{G_F^2|V_{cb}|^2 |{\bf p}\,| q^2}{96\pi^3 m_B^2}\left(1-\frac{m_\tau^2}{q^2}\right)^2\left(H_{--}^2+H_{++}^2+H_{00}^2\right)\,,\nonumber\\
\frac{d\Gamma_\tau}{dq^2}(\lambda_\tau=1/2)&=\frac{G_F^2|V_{cb}|^2 |{\bf p}\,| q^2}{96\pi^3 m_B^2}\left(1-\frac{m_\tau^2}{q^2}\right)^2\frac{m_\tau^2}{2q^2}\left(H_{--}^2+H_{++}^2+H_{00}^2+3 H_{0t}^2\right)\,.
\label{eq:dGamma2}
\end{align}
Again we can define a useful tau spin asymmetry
\begin{equation}
A_{\lambda}(q^2)=\frac{{d\Gamma_\tau}/{dq^2}(\lambda_\tau=-1/2)-{d\Gamma_\tau}/{dq^2}(\lambda_\tau=1/2)}{{d\Gamma_\tau}/{dq^2}}\,,
\end{equation}
which has the explicit form
\begin{equation}
A_{\lambda}(q^2)= 1 - \frac{6|H_{0t}|^2 m_\tau^2}{(2q^2+m_\tau^2)(|H_{--}|^2+|H_{00}|^2+|H_{++}|^2)+3|H_{0t}|^2 m_\tau^2}.
\end{equation}
The presently allowed ranges for this asymmetry are shown in the right plot in Fig~\ref{fig:Asym1}. We observe that also in this observable significant deviations from SM predictions can be expected. Even in the inclusive asymmetry, integrated over $q^2$, where the SM predicts $A_{\lambda,\rm SM}=0.829(15)$, our NP benchmark point yields $A_{\lambda,\rm NP}=0.36$, while even slightly  negative values are still possible.

 \begin{figure}[t]
\centering
\includegraphics[width=0.45\textwidth]{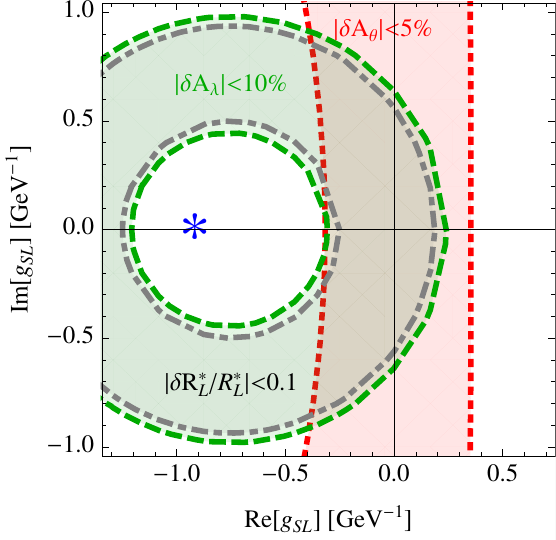}
\caption{Regions allowed by potential future $10\%$ relative precision measurement of $R_L^*$ (shaded in gray and bounded by dot-dashed lines), $10\%$ precision determination of $A_\lambda$ (shaded in green and bounded by dashed lines), and $5\%$ precision measurement of $A_\theta$ (shaded in red and bounded by dotted lines) in the complex plane of NP parameter $g_{SL}$ appearing in the effective $b\to c$ charged current in Eq.~\eqref{eq:Jeff} ($g_{SR}$ is set to zero identically). All observables are assumed to be SM like. The present $B\to D^{(*)} \tau \bar \nu_\tau$ decay rates' best fit NP benchmark point  is marked with the symbol $*$.\label{fig:sens}}
\end{figure}

%
\section{Conclusions}
\label{sec:Conclusions}
%
Within the effective field theory approach we have studied the most general lowest dimensional contributions to helicity suppressed (semi)leptonic $b \to c$ transitions and found that a precise study of the exclusive decay mode $B\to D^* \tau \bar \nu_\tau$ could clarify the possible existence of such non-SM physics. 

Most hadronic inputs entering the theoretical predictions for the $B\to D^* \tau \bar \nu_\tau$ rates can be substantially reduced by normalizing to the $B\to D ^* e \bar \nu_e$ mode.
Using the helicity amplitude formalism we have investigated the presence of $\tau$ mass suppressed helicity amplitudes not observable in $B\to D ^* \ell \bar \nu_\ell$ with $\ell=e$, $\mu$.  We have estimated these contributions using heavy quark expansion including leading perturbative and power corrections and derived precise predictions for the $B\to D^* \tau \bar \nu_\tau$ branching fractions in the SM.  In order to further refine these estimates, Lattice QCD  results for the form factor ratio $A_0/A_1$ (defined in the Appendix~\ref{sec:A2}) would be most valuable.

The $B\to D^* \tau \bar \nu_\tau$ mode has an unique sensitivity to the pseudoscalar density operator which does not contribute to the $B\to D \tau \nu_\tau$ decay mode, while the opposite is true for the scalar density operator. 
Consequently, the precise experimental study of both $B\to D^{(*)} \tau \bar \nu_\tau$ decay modes  can be extremely useful in constraining these kinds of beyond SM physics, especially, since present branching fraction measurements of all these helicity suppressed semileptonic modes are systematically above SM predictions.

Due to the rich spin structure of the $B\to D^* \tau \bar \nu_\tau$ final state one can introduce new observables such as the longitudinal polarization fraction of the $D^*$ ($R_L^*$),  the $D^*-\tau$ opening angle asymmetry ($A_\theta$) and the $\tau$ helicity asymmetry ($A_\lambda$). The discriminating power of these observables is demonstrated in Fig.~\ref{fig:sens} where we show the constraints that such possible future measurements would impose on the NP parameter space for a specific case of $g_{SR}$, for which present $B\to D^{(*)} \tau \bar \nu_\tau$ rate measurements prefer values away from zero (SM). We see that $\mathcal O(10\%)$ precision measurements of $R_L^*, A_{\theta}, A_\lambda$ could critically probe such effects. We have also determined the differential ratios $R^*$, $R^*_L$ and asymmetries $A_\theta$, $A_\lambda$ as a function of the leptons' invariant mass squared. All observables still allow for significant modifications of the corresponding SM predictions and could thus help to disentangle the short distance contributions to $B\to D^{(*)} \tau \bar \nu_\tau$ rates. In term, our study points out the importance of precision measurements of $B \to D^* \tau \bar \nu_\tau$ at the LHCb experiment, as well as at the future Super B factories.   

\begin{acknowledgments}
We acknowledge useful discussions with Bo\v stjan Golob. We are grateful to Manuel Franco Sevilla for pointing out several relevant typos in the first version of the paper.
This work was supported in part  by the Slovenian Research Agency. 
\end{acknowledgments}

\appendix

%
\section{Kinematics and Helicity Structure of the $B \to D^* \tau \bar \nu_\tau$}
\label{sec:A1}
%

Following~\cite{Korner-Schuler1} we consider the kinematics of the decay of a $B$ meson to a polarized $D^*$ together with a charged lepton - antineutrino pair. 
As throughout the paper, we will assume that the decay is mediated by interactions of the form \eqref{eq:Heff}, i.e. that the leptons are produced via the standard left-handed charged current as in the SM, while NP could modify quark charged currents. However, generalization of our results to non-standard forms of the leptonic charged currents is straightforward. We denote the momenta of $B, D^*, \ell, \nu$ with $p_B, p_{D^*}, k_\ell, k_\nu$ respectively, while $q\equiv p_B-p_{D^*}=k_\ell+k_\nu$. We also introduce the angles $\theta$ between the $D^*$ and $\tau$ three-momenta in the $\tau-\bar\nu_\tau$ rest frame, as well as $\chi$, between the plane of the charged lepton and antineutrino momenta, and the  decay plane of the $D^*$. Helicity basis vectors of the $D^*$ (vector) meson are denoted as $\varepsilon^{\alpha}$, while assuming standard lepton interactions, we can use $\tilde{\varepsilon}^\mu$ for the four basis vectors describing the total helicity of the charged lepton-neutrino system.  
In the rest frame of the $B$ meson with $z$ axis along the trajectory of the $D^*$, a suitable basis for the lepton pair helicities is
\begin{align}
\tilde{\varepsilon}_\mu(\pm)&=\frac{1}{\sqrt{2}}(0,\pm1,-i,0)\,, \nonumber\\
\tilde{\varepsilon}_\mu(0)&=\frac{1}{\sqrt{q^2}}(|{\bf p}\,|,0,0,-q_0)\,,\nonumber\\
\tilde{\varepsilon}_\mu(t)&=\frac{1}{\sqrt{q^2}}(q_0,0,0,-|{\bf p}\,|),
\end{align}
where $q_0=(m_B^2-m_{D^*}^2+q^2)/{2m_B}$ and 
\beq
|{\bf p}\,|=\frac{\lambda^{1/2}(m_B^2,m_{D^*}^2,q^2)}{2m_B}\,, 
\label{eq:pD}
\eeq
with $\lambda(a,b,c) = a^2+b^2+c^2 - 2(ab+bc+ca)$. They satisfy the following normalization and completeness relations
\begin{equation}
\tilde{\varepsilon}_\mu^*(m)\tilde{\varepsilon}^\mu(m')=g_{mm'},\quad {\rm for}\quad (m,m'=t,\pm,0)\,,
\end{equation}
\begin{equation}
\sum_{m,m'}\tilde{\varepsilon}_\mu(m)\tilde{\varepsilon}_\nu^*(m')g_{mm'}=g_{\mu\nu}\,.
\end{equation}
Similarly a convenient helicity basis for the $D^*$ is
\begin{align}
\varepsilon_\alpha(\pm)&=\mp\frac{1}{\sqrt{2}}(0,1,\pm i,0)\,, \nonumber\\
\varepsilon_\alpha(0)&=\frac{1}{m_{D^*}}(|{\bf p}\,|,0,0,E_{D*}),
\end{align}
where $E_{D^*}=(m_B^2+m_{D^*}^2-q^2)/2m_B$ is the energy of $D^*$ in the $B$ rest frame.
These basis vectors satisfy the following normalization
\begin{equation}
\varepsilon_\alpha^*(m)\varepsilon^\alpha(m')=-\delta_{mm'}\,,
\end{equation}
and completeness relation
\begin{equation}
\sum_{mm'}\varepsilon_\alpha(m)\varepsilon_\beta(m')\delta_{mm'}=-g_{\alpha\beta}+\frac{p_{D^*\alpha}p_{D^*\beta}}{m_{D^*}^{2}}.
\end{equation}
We can now introduce helicity amplitudes, $H_{\pm\pm}$, $H_{00}$ and $H_{0t}$ describing the decay of a pseudo-scalar meson into the three helicity states of a vector meson and four helicity states of the leptonic pair
\begin{align}
H_{mm}(q^2)&=\tilde{\varepsilon}(m)^{\mu*}H_{\mu}(m),\quad {\rm for} \quad m=0,\pm\,, \nonumber \\
H_{0t}(q^2)&=\tilde{\varepsilon}(m=t)^{\mu*}H_{\mu}(n=0).
\end{align}
Here, $H_\mu(m)$ is a corresponding hadronic matrix element, and $m,n$ denote helicity projections of the $D^*$ meson and the leptonic pair in the $B$ rest frame. 

If mediated by the effective Hamiltonian of the form~\eqref{eq:Heff} for arbitrary quark charged current $J_{bc}^\mu$, the $B\to D^* \ell \nu_\ell$ triply differential decay rate can be written as
\begin{equation}
\frac{d^2\Gamma_\ell}{dq^2d \cos\theta d\chi}=\frac{G_F^2 |V_{cb}|^2}{(2\pi)^4}\frac{|{\bf p}\,|}{2m_B^2}\left(1-\frac{m_\ell^2}{q^2}\right)  L_{\mu\nu}H^{\mu\nu},
\end{equation}
where $L_{\mu\nu}$,$H_{\mu\nu}$ are the leptonic and hadronic current tensors. Using completeness relations of the helicity basis vectors we can  rewrite $L_{\mu\nu}H^{\mu\nu}$ as
\begin{equation}
L_{\mu\nu}H^{\mu\nu}=L_{\mu'\nu'}g^{\mu'\mu}g^{\nu'\nu}H_{\mu\nu}=\sum_{mm',nn'}\Big(L_{\mu'\nu'}\tilde{\varepsilon}^{\mu'}(m)\tilde{\varepsilon}^{\nu'}(n)g_{mm'}g_{nn'}\Big)\Big(\tilde{\varepsilon}^{\mu\ast}(m')\tilde{\varepsilon}^{\nu}(n')H_{\mu\nu}\Big).
\end{equation}
Following~\cite{Korner-Schuler2,Kadeer-Schuler} we can expand the leptonic tensor in terms of a complete set of Wigner's $d^J$ functions, reducing $L_{\mu\nu}H^{\mu\nu}$ to the following compact form
\begin{equation}
\begin{split}
L_{\mu\nu}H^{\mu\nu}=\frac{1}{8}\sum_{{\small{\lambda_\ell, \lambda_{D*}, \lambda_{\ell\nu}, \lambda'_{\ell\nu}, J, J'}}} (-1)^{J+J'} |h_{(\lambda_\ell, \lambda_\nu)}\mid^2\delta_{\lambda_{D*}\lambda_{\ell\nu}}\delta_{\lambda_{D*}\lambda'_{\ell\nu}}\\
\times d^J_{\lambda_{\ell\nu}, \lambda_{\ell}-1/2}(\theta)d^J_{\lambda'_{\ell\nu}, \lambda_{\ell}-1/2}(\theta)H_{\lambda_{D*}\lambda_{\ell\nu}}H^*_{\lambda_{D*}\lambda'_{\ell\nu}}\,,
\end{split}
\end{equation}
where $J$ and $J'$ run over $1$ and $0$.  In term, the lepton helicity amplitudes, $h_{(\lambda_\ell,\lambda_\nu)}$ for a left-handed weak current are given by
\begin{equation}
h_{(\lambda_\ell,\lambda_\nu)}=\frac{1}{2} \bar{u}_\ell({\lambda_\ell})\gamma^\mu(1-\gamma^5)v_{\nu}(\lambda_\nu)\tilde{\epsilon}_{\mu}(\lambda_{\ell\nu}),
\end{equation}
where for massless right-handed antineutrinos $\lambda_\nu=1/2$ and  $\lambda_{\ell\nu}=\lambda_\ell-\lambda_\nu$ in the $\ell{\nu}$ center of mass frame by angular momentum conservation. It follows that the two non-vanishing $|h_{(\lambda_\ell, \lambda_\nu)}\mid^2$ contributions are
\begin{eqnarray}
\vert h_{-1/2,1/2}\vert^2=2(q^2-m_\ell^2)\, \quad {\rm and} \quad \vert h_{1/2,1/2}\vert^2=2\frac{m_\ell^2}{2q^2}(q^2-m_\ell^2)\,.
\end{eqnarray}
Finally, using the standard~\cite{PDG} convention for Wigner's d-functions and performing the trivial integration over $\chi$ we obtain 
\begin{align}
\frac{d^2\Gamma_\ell}{dq^2 d \cos\theta}(\lambda_\ell=-1/2) &= \frac{G_F^2 |V_{cb}|^2|{\bf p}\,|q^2}{256\pi^3 m_B^2} \left(1-\frac{m_\ell^2}{q^2}\right)^2 \left[(1-\cos\theta)^2 H_{++}^2+(1+\cos\theta)^2 H_{--} ^2+2\sin^2\theta H_{00}^2\right]\,,\nonumber\\
\frac{d^2\Gamma_\ell}{dq^2 d \cos\theta}(\lambda_\ell=1/2) &= \frac{G_F^2 |V_{cb}|^2|{\bf p}\,|q^2}{256\pi^3 m_B^2} \left(1-\frac{m_\ell^2}{q^2}\right)^2 \frac{m_\ell^2}{q^2}\left[(\sin^2\theta(H_{++}^2+H_{--}^2)+2(H_{0t}-H_{00}\cos\theta)^2\right]\,,\label{eq:Gammaf}
\end{align}
from which Eqs. \eqref{eq:dGamma}, \eqref{eq:dGamma1} and \eqref{eq:dGamma2} can be easily derived via summation over $\lambda_\ell$ and/or integration over $\cos\theta$.

%
\section{Helicity amplitudes and hadronic matrix elements}
\label{sec:A2}
%

In the SM (in presence of only the first term in~\eqref{eq:Jeff}), the helicity amplitudes $H_{mn}$ can be writen as
\begin{align}
H^{\rm SM}_{\pm\pm}(q^2)&=(m_B+m_{D^*})A_1(q^2)\mp\frac{2m_B}{m_B+m_{D^*}}|{\bf p}\,|V(q^2)\,,\nonumber\\
H^{\rm SM}_{00}(q^2)&=\frac{1}{2m_{D^*}\sqrt{q^2}}\left[(m_B^2-m_{D^*}^2-q^2)(m_B+m_{D^*})A_1(q^2)-\frac{4m_B^2|{\bf p}\,|^2}{m_B+m_{D^*}}A_2 (q^2)\right]\,,\nonumber\\
H^{\rm SM}_{0t}(q^2)&=\frac{2m_B|{\bf p}\,|}{\sqrt{q^2}}A_0(q^2)\,.
\end{align}
where the form factors parametrizing the relevant hadronic matrix elements are defined as
\begin{subequations}
\begin{eqnarray}
\langle D^*(p_{D^*},\epsilon_\alpha)|\bar{c}\gamma_\mu b|{B(p_B)}\rangle
&=& \frac{2 i V(q^2)}{m_B+m_{D^*}}\;\epsilon_{\mu\nu\alpha\beta}
\epsilon^{*\nu}p_B^\alpha p_{D^*}^\beta \,,
\label{one}\\
\langle D^*(p_{D^*},\epsilon_\alpha)|\bar{c}\gamma_\mu\gamma_5 b|{B(p_B)}\rangle
&=& 2m_{D^*}\,A_0(q^2)\frac{\epsilon^*\cdot q}{q^2}q_\mu
+ (m_B+m_{D^*})\,A_1(q^2)\,\left(\epsilon^*_\mu - \frac{\epsilon^*\cdot q}{q^2}q_\mu
\right) \nonumber\\
& &- A_2(q^2)\,\frac{\epsilon^*\cdot q}{m_B+m_{D^*}}\left( (p_B+p_{D^*})_\mu 
-\frac{m_B^2-m_{D^*}^2}{q^2}q_\mu\right)\,.
\label{two}
\end{eqnarray}
\end{subequations}

In presence of NP of the form~\eqref{eq:Jeff}, one needs to evaluate two additional matrix elements given by
\begin{subequations}
\begin{eqnarray}
\langle D^*(p_{D^*},\epsilon_\alpha)|\bar{c} b|{B(p_B)}\rangle &=& 0\,,\\
\langle D^*(p_{D^*},\epsilon_\alpha)|\bar{c} \gamma_5b|{B(p_B)}\rangle&=&\frac{1}{m_b+m_c}q^\mu\langle D^*|\bar{c}\gamma_\mu\gamma_5 b|\bar{B}^0\rangle = \frac{2m_{D^*}}{m_b+m_c} \,A_0(q^2){\epsilon^*\cdot q} \,.
\end{eqnarray}
\end{subequations}
The final effect of such contributions to the differential rates can be encoded into the $H_{0t}$ helicity amplitude as given in Eq.~\eqref{eq:HNP}. 

We can use further information on the form factors given by precise differential decay rate measurements in $B\to D^*\ell\bar \nu_\ell$~\cite{Belle}, as well as their perturbatively computable properties and relations in the heavy quark limit for the $b,c$ quarks~\cite{Falk:1992wt, Neubert}. In this limit it is customary to employ a new kinematical variable
\begin{equation}
 w \equiv v_B\cdot v_{D^*}= \frac{m_B^2 + m_{D*}^2-q^2}{2 m_B {m_{D*}}},
\end{equation}
with $v_B^\mu$, and $v_{D^*}^\mu$ being the four-velocities of the $B$ and $D^*$ meson respectively. One can then define an universal form factor 
\beq
h_{A_1} (w) =  A_1(q^2) \frac{1}{R_{D^*}}\frac{2}{w+1}\,,
\eeq
and ratios $R_1$, $R_2$ and $R_0$ in terms of which
\begin{align}
  A_0(q^2)&=\frac{R_0(w)}{R_{D^*}^{}}h_{A_1}(w)\,,\nonumber\\
  A_2(q^2)&=\frac{R_2(w)}{R_{D^*}^{}}h_{A_1}(w)\,,\nonumber\\
  V(q^2)&=\frac{R_1(w)}{R_{D^*}^{}}h_{A_1}(w)\,,
\end{align}
where $R_{D^*}=2\sqrt{m_Bm_{D^*}}/(m_B+m_{D^*})$. 
The $w$ dependence of these quantities in the heavy quark limit reads
\begin{eqnarray}
  h_{A_1}(w) & = &
  h_{A_1}(1)\big[1-8\rho^2z+(53\rho^2-15)z^2-(231\rho^2-91)z^3\big]~, \nonumber\\
  R_1(w) & = & R_1(1)-0.12(w-1)+0.05(w-1)^2,\nonumber\\ 
  R_2(w) & = & R_2(1)+0.11(w-1)-0.06(w-1)^{2},\nonumber\\
  R_0(w) & = & R_0(1)-0.11(w-1)+0.01(w-1)^{2},
\end{eqnarray}
where $z=(\sqrt{w+1}-\sqrt{2})/(\sqrt{w+1}+\sqrt{2})$. The first three expressions can be found in~\cite{Caprini-Neubert}, while we have derived the fourth using the results of~\cite{Caprini-Neubert}.
Above relations contain free parameters $h_{A_1}(1),R_1(1),R_2(1), \rho^2$, which can be extracted from the well measured $B\to D^*\ell\bar \nu_\ell$ decay distributions.  In our numerical evaluation of the $B\to D^*\tau\bar \nu_\tau$ differential decay rates we employ the results of a recent Belle analysis~\cite{Belle}. The virtue of this approach is that most of the associated hadronic uncertainties actually cancel in ratios of decay rates to tau versus light leptons, as previously demonstrated for the case of $B\to D \tau \bar \nu_\tau$~\cite{Kamenik:2008tj}. 

In addition to these inputs, the $B\to D^* \tau \bar \nu_\tau$ rate also depends on $R_0(1)$, which cannot be extracted from $B\to D^*\ell\bar \nu_\ell$ studies, since it only appears in the helicity suppressed amplitude $H_{0t}$. In the exact heavy quark limit $[R_0(1)]_{\rm HQET} = 1$. Leading order perturbative (in $\alpha_s$) and power ($1/m_{b,c}$) corrections are known~\cite{Falk:1992wt, Neubert} for the linear combination
\beq
R_3(1) \equiv \frac{R_2(1)(1-r) + r [R_0(1) (1+r) -2] }{(1-r)^2} = 0.97\,.
\label{eq:R3}
\eeq
The same calculation predicts $R_2(1)=0.80$, yielding eventually $R_0(1) = 1.22$\,. Experimentally ~\cite{Belle} however,  $R_2(1) = 0.864(25)$ and inserting this value into Eq.~\eqref{eq:R3} yields our final result $R_0(1) = 1.14$. In all our numerical calculations we use directly Eq.~\eqref{eq:R3} and conservatively assign a $10\%$ uncertainty to this value accounting for higher order corrections. In the future, a more reliable and precise determination of $R_0(1)$ (or equivalently $R_3(1)$) could be obtained on the lattice, similarly as has already been done for the helicity suppressed $B\to D$ matrix elements~\cite{Tantalo:2007uy}.

\end{document}